# High throughput data-driven design of laser crystallized 2D MoS$_2$ chemical sensors

Drake Austin[1,2], Paige Miesle[1,2,3], Deanna Sessions[1,2,4], Michael Motala[1,2], David Moore[1,2], Griffin Beyer[1,2,3], Adam Miesle[1,2,3], Andrew Sarangan[3], Amritanand Sebastian[5], Saptarshi Das[5,6], Anand Puthirath[7], Xiang Zhang[8], Jordan Hachtel[8], Pulickel Ajayan[7], Tyson Back[1], Peter Stevenson[1], Michael Brothers[2,9], Steven Kim[9], Philip Buskohl[1], Rahul Rao[1], Christopher Muratore[3*], and Nicholas Glavin[1*]

[1]Air Force Research Laboratory, Materials and Manufacturing Directorate, WPAFB, OH, USA
[2]UES Inc., Dayton, OH, USA
[3]University of Dayton, OH, USA
[4]Department of Electrical Engineering, Pennsylvania State University, University Park, PA, USA
[5]Department of Engineering Science and Mechanics, Pennsylvania State University, University Park, PA, USA
[6]Department of Materials Science and Engineering, Pennsylvania State University, University Park, PA, USA
[7]Rice University, Houston, TX, USA
[8]Center for Nanophase Materials Science, Oak Ridge National Laboratory, Oak Ridge, TN, USA
[9]Air Force Research Laboratory, 711th Human Performance Wing, WPAFB, OH, USA

*Corresponding author.
E-mail address: cmuratore1@udayton.edu, nicholas.glavin.1@us.af.mil

Abstract: High throughput characterization and processing techniques are becoming increasingly necessary to navigate multivariable, data-driven design challenges for sensors and electronic devices. For two-dimensional materials, device performance is highly dependent upon a vast array of material properties including number of layers, lattice strain, carrier concentration, defect density, and grain structure. In this work, laser-crystallization was used to locally pattern and transform hundreds of regions of amorphous MoS$_2$ thin films into 2D 2H-MoS$_2$. A high throughput Raman spectroscopy approach was subsequently used to assess the process-dependent structural and compositional variations for each illuminated region, yielding over 5500 distinct non-resonant, resonant, and polarized Raman spectra. The rapid generation of a comprehensive library of structural and compositional data elucidated important trends between structure-property-processing relationships involving laser-crystallized MoS$_2$, including the relationships between grain size, grain orientation, and intrinsic strain. Moreover, extensive analysis of structure/property relationships allowed for intelligent design, and evaluation of major contributions to, device performance in MoS$_2$ chemical sensors. In particular, it is found that sensor performance is strongly dependent on the orientation of the MoS$_2$ grains relative to the crystal plane.

High-throughput, data-driven design of nanomaterials is critical to ensure the rapid development of devices that operate across application areas including electronics, sensing, and optoelectronics. Relevant to sensor applications, two-dimensional (2D) transition metal dichalcogenides (TMDCs) exhibit exceptional electrical, optical, mechanical, and chemical properties with the benefit of having a high surface to volume ratio. Dependent on the chemical composition and phase configuration[1], TMDCs can present as semiconductors, semi-metals, and metals[2-4], and are comprised of covalently-bonded layer sheets held together by weak van der Waals forces in the bulk form. The electrical and optical properties of TMDCs can be tuned by varying the number of layers[5-8], applying strain[9-14], and introducing dopants or defects[15-20]. Moreover, the high surface to volume ratio, high electron mobility, and ease of functionalization make TMDC materials an ideal platform for electronic detection of a multitude of liquid and vapor analytes[21].

Data-driven design of electronic sensors is an attractive approach to efficiently and rapidly correlate multibody intrinsic material properties relevant to sensor performance optimization[22]. In the case of 2D $MoS_2$ chemical sensors, detection sensitivity is known to be non-monotonically responsive to the layer thickness[23], grain size[24-26], and defect density[27,28]. Moreover, 2D sensor devices are often fabricated from a finite batch of samples grown by wafer-scale, relatively uniform techniques including thermal decomposition, metal-organic chemical vapor deposition, or magnetron sputtering[29-31]. Following the growth process, annealing strategies such as thermal or photonic annealing can tailor the crystallinity, induce defects, or even manipulate electronic properties[19,32,33]. For films produced by such wafer-scale techniques, the tuning of these material properties must be done on a per-sample basis or must rely on film variation intrinsically present[34]. As a result, evaluating the relationships between the varying material properties is a time-consuming process.

By introducing strategies to locally modify 2D materials after film growth, a large parameter space for material tuning can be achieved. For instance, exposure to ion beams has proven a viable technique to controllably introduce defects in TMDC films, allowing for a determination of the relationships between induced defects and the measured Raman and photoluminescence spectra[16]. Alternatively, laser patterning has been used to produce localized patterns within TMDC films through thermal decomposition[35] or thermally-induced chemical reactions[36], induce crystallization[37-39], introduce or remove dopants and defects[40-42], and induce structural changes[43-45]. However, the potential large-scale material variation associated with such techniques has revealed minimal connections to actual device applications such as chemical sensors discussed in this work.

To investigate the influence of such structure-property-processing relationships on sensing performance, wafer-scale few-layer films of amorphous $MoS_2$ ($a$-$MoS_2$) were grown using magnetron sputtering, and select localized regions were crystallized to 2H-$MoS_2$ through laser exposure. This investigation was broken into two parts. First, multiple samples with many laser-crystallized regions were fabricated for the purpose of characterization across a wide range of laser-processing conditions and their corresponding material properties. As the laser intensity and exposure time can be simply controlled, hundreds of micro- to millimeter scale regions of crystallized $MoS_2$ can be patterned on a single sample at high throughput, each with varying material properties. From these conditions, thousands of individually unique Raman spectra across multiple samples were characterized to determine the relationship between strain, doping, out-of-plane orientation, defect density, and defect relationships as a function of our laser crystallization approach. Second: based on the structure-property-processing relationships uncovered from the aforementioned characterization, a smaller set of laser-processing conditions and material properties were used in the fabrication of $MoS_2$-based sensors. From this high-throughput and data-driven approach, the interplay between intrinsic material properties and sensor device performance were evaluated. As such, critical design parameters were revealed and discussed below for $MoS_2$-based sensors across different film thicknesses.

## Results and Discussion

### Laser-induced crystallization of $MoS_2$ thin-films

Films of $a$-$MoS_2$ were grown on 0.5 mm thick Corning Eagle Glass substrates by pulsed DC magnetron sputtering at room temperature[32,46]. The magnetron sputtering technique is capable of producing samples with film thicknesses ranging from as few as four atomic layers to bulk films[47]. In this first part of the study, six samples were fabricated with film thicknesses of 3.2, 3.6, 4.9, 6.5, and 7.8 nm. These amorphous samples were then exposed to a scanned, continuous-wave (CW) laser beam with a central wavelength of 514.5 nm. Laser exposure occurred while samples were immersed in an argon environment (Figure 1(a)) to reduce reactions with ambient oxygen and moisture[48]. To produce uniform sample areas suitable for spectroscopic and electrical interrogation, the focused beam was raster-scanned eight times across the surface of the sample with a raster separation of 6 μm (center-to-center) over a length of 1 mm. Example confocal microscope images of the resulting laser-written lines are shown in Figure 1(b) (see Figure S1 for the full image). The laser was scanned over the surface with translation speeds of v = 0.1, 0.3, and 1.0 mm/s and the range of laser power was selected from $P$ = 25–300 mW. Independent control of these parameters resulted in localized control of the heating time and the peak temperature reached. In the case of CW lasers, the peak temperature is determined by the linear power density $P_{lin} = P/w$ independent of the focal spot size[49] where w is the $1/e^2$ focal spot radius.

In this first part of the study, seven samples were fabricated with film thicknesses of 2.4, 3.2, 3.6, 4.9, 6.5, and 7.8 nm. However, the 2.4 nm thick sample was omitted from this analysis due to issues with sample quality.

### High-throughput characterization of laser-crystallized $MoS_2$

Raman spectroscopy was used as the primary method of characterization as its high spatial resolution (1 μm) allowed for the acquisition of many spectra within every laser-crystallized region. A total of 4002 unpolarized Raman spectra using a 514.5 nm excitation laser were acquired within localized crystalline regions across samples of six different film thicknesses (3.2, 3.6, 4.9, 6.5, and 7.8 nm). This large number of spectra was acquired by scanning the Raman laser over the width of each laser-written line, resolving the spatial variation in the Raman signal due to the annealing laser's Gaussian focal spot profile (Figure S2).

The values of the peak intensity $I_i$, frequency $\omega_i$, and linewidth $\Gamma_i$ (as determined by the full width at half maximum (FWHM)) of the $E_{2g}$ and $A_{1g}$ peaks ($i = E, A$, respectively) were extracted for each spectrum by fitting to two pseudo-Voigt functions with the initial guesses determined automatically using peak-finding software (additional details found in supplementary information including example fits in Figure S3). The data was filtered to remove spectra with $E_{2g}$ peak intensities in the bottom 5th percentile of each laser-written line. This was done in order to exclude particularly weak Raman signals, i.e. the peripheries of the laser-written lines. Examples of the normalized spectra are shown in a waterfall plot in Figure 1(c) where the stacking order is determined by the extracted $E_{2g}$ frequency and the color scale is determined by the extracted $E_{2g}$ FWHM linewidth.

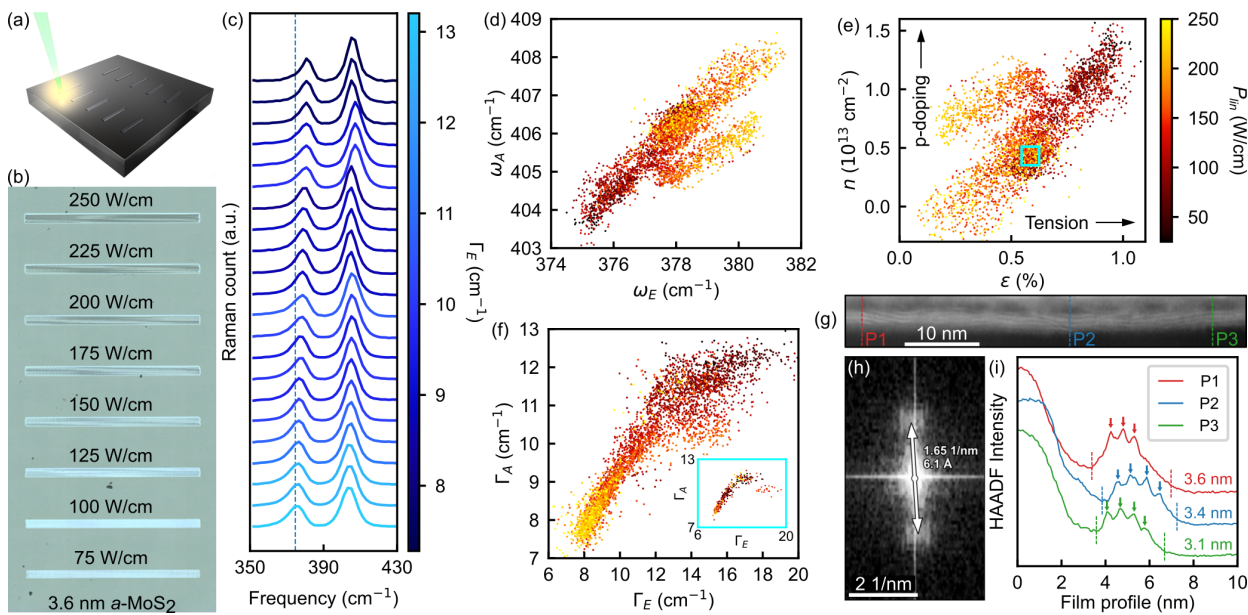

Figure 1: *High-throughput analysis of Raman spectra in laser-crystallized 2H-MoS$_2$. (a) Schematic of laser processing with a 514.5 nm laser. (b) Optical image of laser-crystallized lines. (c) Waterfall plot of select 514.5 nm Raman spectra descending in order of decreasing $E_{2g}$ frequency. The color scale indicates the $E_{2g}$ FWHM. (d) Relationship between the $A_{1g}$ and $E_{2g}$ frequencies. The color scale indicates the linear power density used for annealing. The smaller cluster of linear data with reduced slope corresponds to data points from the sample with the thinnest pre-annealed film (3.2 nm). (e) Strain and doping associated with the $E_{2g}$ and $A_{1g}$ frequencies. (f) Relationship between the $A_{1g}$ and $E_{2g}$ linewidths. A linear trend is observed for narrower linewidths that becomes non-linear for broader linewidths. Inset: Same plot, but with the data set restricted to the strain and doping values indicated by the cyan box in (e). (g) STEM image of a 3.2 nm thick region annealed with $P_{lin}$ = 125 W/cm and v = 1.0 mm/s. (h) FFT of image in (g) showing interlayer spacing. (i) Plots corresponding to the profiles indicated in (g).*

Using this data, the key relationships observed between the extracted fitting parameters and the physical characteristics with which they correspond can be analyzed across all film thicknesses. The extracted frequencies of the $A_{1g}$ and $E_{2g}$ Raman peaks ($\omega_A$ and $\omega_E$) of over 4000 unique spectra are shown in Figure 1(d), where the color scale indicates the linear power density of the annealing laser ($P_{lin}$). Two linear trends with distinctive slopes are observed, with higher values of $P_{lin}$ tending to result in crystallites with higher peak frequencies for both $E_{2g}$ and $A_{1g}$ Raman modes. The linear region with the shallower slope corresponds specifically to the thinnest amorphous film with a pre-annealed thickness of 3.2 nm, suggesting that the mechanisms responsible for the frequency shifts are influenced by the number of molecular layers for thinner films. While film thickness can influence peak frequencies by red-shifting the $A_{1g}$ peak and blue-shifting the $E_{2g}$ peak with increasing thickness[7,9], the frequencies will be within 0.5 cm$^{-1}$ of bulk values when the film thickness reaches 5 molecular layers[8]. This suggests that variations in the film thickness alone cannot account for the observed frequency shifts. However, strain and doping are also known to influence the frequencies of these peaks[9,10,34,50,51], with strain in the lateral plane primarily influencing $\omega_E$ and doping primarily influencing $\omega_A$. While the samples here were not externally strained during analysis, internal in-plane and out-of-plane strain may still be present due to lattice mismatches between grains. The frequency shift $\Delta\omega$ can be related to strain and doping within the material through the following expression[50]:

$$\Delta\omega_i = -2\gamma_i\omega_i^0\epsilon + k_{n,i}n \tag{1}$$

where $\gamma$ is the Grüneisen parameter, $\omega^0$ is the frequency corresponding to undoped and unstrained MoS$_2$, $\epsilon$ is the amount of strain, $k_n$ is an empirical constant, $n$ is the carrier concentration (negative for electrons, positive for holes), and the subscript $i = E, A$ indicates that the value corresponds to either the E$_{2g}$ or A$_{1g}$ peak. The first term describes how changes in the volume of the lattice influence its in-plane and out-of-plane vibrational properties (the E$_{2g}$ and A$_{1g}$ Raman modes, respectively). The second term describes an empirically observed, quasi-linear relationship for MoS$_2$[51]. While the values for $\omega^0$ are dependent on film thickness, they can be approximated by the bulk values (382 and 408 cm$^{-1}$ for the E$_{2g}$ and A$_{1g}$ peaks, respectively), as previously discussed. However, bulk values for $\gamma$ and $k_n$ have not been reported to the best of the authors' knowledge. Using the monolayer MoS$_2$ values for $\gamma$ and $k_n$ instead[50], Eq. 1 provides a one-to-one mapping between the peak frequencies in Figure 1(d) to the strain and doping, as shown in Figure 1(e). Overall, it is found that the primary charge carriers are holes and the crystal grains are under tension. As few-layer films were used in this study, the appropriate values for $\gamma$ and $k_n$ may differ from those of monolayer films, and so the absolute values of the strain and carrier concentration are only an estimate of quantitative values. However, the relative differences between data points can still be compared. Again, two linear trends are observed, with the region with the smaller slope corresponding to the 3.2 nm film (see Figure S5 for the same plot with the 3.2 nm film data excluded). Equation (1) provides a plausible explanation for the presence of this region: the values of $\gamma$, $k_n$, and $\omega^0$, are dependent on the number of molecular layers up until approximately 4 layers. The values of $\gamma$ and $k_n$ dictate the slope of the strain vs. doping plot, while changes in $\omega^0$ result in a shift in the absolute values. For 5 or more molecular layers, the film is bulk-like when considering these values, and the frequency shifts fall on the same line. In either region, however, it is found that less-doped regions tend to be under less strain, with the least amount of doping and strain corresponding to the regions annealed with the largest laser intensity corresponding to the highest localized temperatures.

The influence of annealing temperature on strain can be further analyzed by considering the linewidth of the Raman peaks for MoS$_2$[52] ($\Gamma_A$ and $\Gamma_E$), shown in Figure 1(f). Similar to Figure 1(e), which relates strain to doping, a roughly linear trend is observed when comparing the linewidths $\Gamma_A$ and $\Gamma_E$. This is expected as the E$_{2g}$ and A$_{1g}$ peak linewidths correspond to disorder (i.e. grain size) in the lateral and transverse directions, respectively[53], with smaller linewidths corresponding to less disorder (larger grains). It can be seen that higher annealing temperatures tend to result in larger grains, which offers an explanation for the observed trend in the strain. With increasing lateral disorder, each individual crystallite exerts a greater force on its neighbors at the grain boundary due to the mismatch in their crystallographic order. This results in more overall strain, which is reduced as the grain size increases and the number of grain boundaries decreases[54]. At the largest linewidths, a deviation from linearity is observed, possibly due to changes in doping and strain[50]. This can be examined by restricting the data set to a small range of nearly constant strain and doping values, indicated by the data points within the cyan box in Figure 1(e). This range of values was selected by selecting the highest-populated bin within the strain and doping data. The resulting plot of this restricted data set is shown in the inset of Figure 1(f). The linear trend is still observed for smaller linewidths, but eventually levels off for the $A_{1g}$ peak, suggesting that strain and doping are not the cause of the observed non-linearity. Instead, it is proposed here that, while the amount of lateral disorder ($\Gamma_E$) can continue to increase, the amount of transverse disorder ($\Gamma_A$) is limited by the finite film thickness. To observe the extent of atomic disorder for a typical annealed region, scanning transmission electron microscopy (STEM) was used to image the 3.2 nm thick sample annealed at conditions favorable for high crystalline order ($\Gamma_E$ =11.2 ± 1.7 cm$^{-1}$, $\Gamma_A$ =10.3 ± 0.4 cm$^{-1}$). The resulting STEM image, its fast Fourier transform (FFT), and select lineouts are shown in Figure 1(g-i). The annealed film is 3–4 molecular layers thick (~3.4 nm) with a grain size of roughly 10 nm with most of the crystallites oriented in-plane.

In order to investigate the interplay between strain/doping and crystal grain size, the $E_{2g}$ frequency and linewidth values extracted from the library of spectra collected from the samples are shown in Figure 2(a). In this case, a linear trend is observed for $\Gamma_E$ values below approximately 10 cm$^{-1}$. Beyond this, the $\omega_E$ values level off while also broadening significantly. In order to further elucidate this trend, the data points were colored based on the values of the $A_{1g}/E_{2g}$ peak intensity ratio ($I_A/I_E$) from the unpolarized spectra. The linear and non-linear regions of the plot are visibly separated based on these values, which are known to relate to the Raman excitation laser polarization and the grain orientation[7,55].

Utilizing the large available data set, further analysis was performed using Uniform Manifold Approximation and Projection (UMAP), a dimension reduction technique useful for visualizing multi-dimensional data sets[56]. UMAP is a nonlinear reduction technique that preserves local and global data structure based on pairwise probabilities. This probabilistic approach differs from linear reduction techniques, such as Principal Component Analysis (PCA)[57,58], which are better suited for multi-colinear high dimensional data as they reduce data through a linear change of basis. While PCA is suited for capturing variance in the global data through linear transforms, UMAP is well-suited for identifying continuous trends by preserving global relationships while also maintaining neighborhoods of similar data – see supporting information for a detailed mathematical description of the technique. Applying UMAP to the data set of non-resonant Raman spectra within a frequency range of 325–466 cm$^{-1}$ (i.e. a range containing the $E_{2g}$ and $A_{1g}$ peaks) results in Figure 2(b). Within this plot, each data point represents a single Raman spectrum, as indicated by the inset. The horizontal and vertical axes represent the reduced dimensions onto which the high-dimensional data set has been projected, and do not have a simple physical interpretation. However, distances between points within this plot do have an intuitive interpretation: they represent how similar each point (Raman spectrum) is probabilistically, i.e. across all relevant features of the spectra (see Figure S8 for an illustration of this). By coloring each point based on the value of the corresponding fitting parameters and relationships between them (e.g. $I_A/I_E$ in Figure 2(b)), potential trends can be identified. Here, Figure 2(b) shows significant clustering based on $I_A/I_E$, indicating that points with similar $I_A/I_E$ ratios tend to also be similar in other aspects of their Raman spectra. This overall suggests a potential relationship between grain orientation and other critical variables.

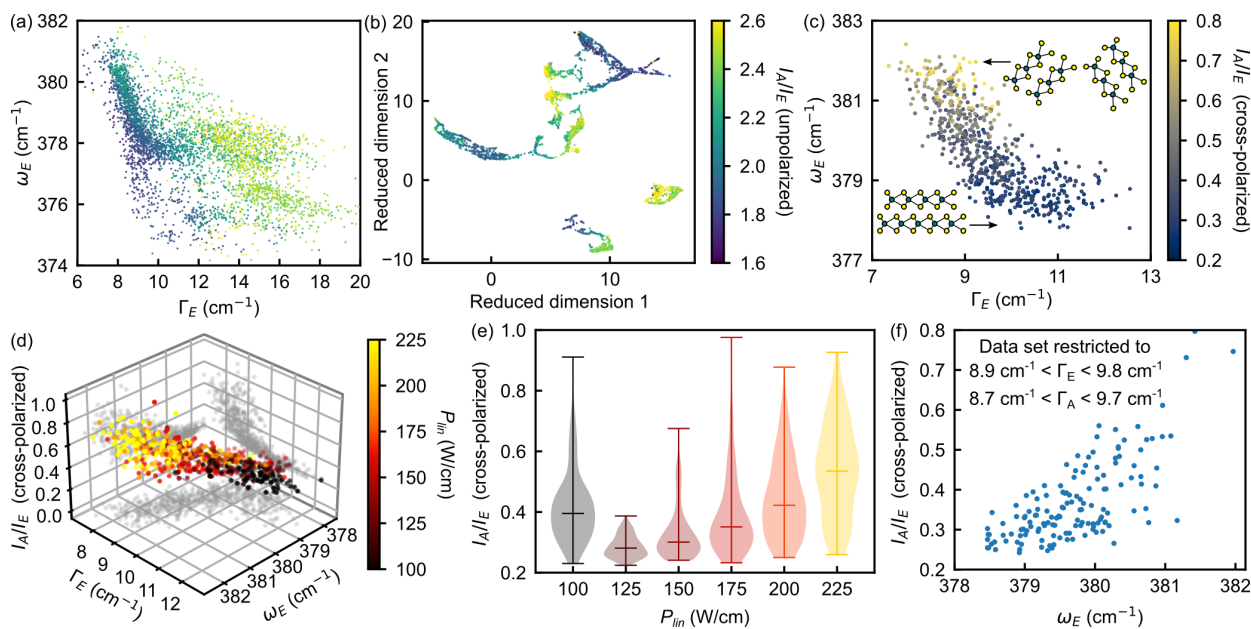

*Figure 2: Analysis of polarized Raman spectra. (a) Relationship between the $E_{2g}$ frequency and the $E_{2g}$ FWHM linewidth with the color scale indicating the $A_{1g}/E_{2g}$ peak intensity ratio. A linear region is observed for narrower linewidths, which also corresponds to points with smaller $A_{1g}/E_{2g}$ ratios. (b) UMAP plot representing a dimensional reduction of all 514.5 nm Raman spectra into a 2D plane (see SI for details). Each point represents a particular Raman spectrum. The color scale indicates the $A_{1g}/E_{2g}$ peak intensity ratio. (c) Same as (a), but under 514.5 nm cross-polarized Raman spectroscopy for the 3.4 nm film. Higher frequencies are observed, likely due to oxidation of the sample since its original characterization under unpolarized Raman spectroscopy. Inset images illustrate grain orientation for high and low $A_{1g}/E_{2g}$ peak intensity ratios. (d) 3D plot with 2D projections showing the relationships between the $A_{1g}/E_{2g}$ peak intensity ratio and the $E_{2g}$ FWHM and frequency. The color scale indicates the linear power density. (e) Violin plot showing the distribution of $A_{1g}/E_{2g}$ peak intensity ratios for each linear power density used. (f) Plot of the $A_{1g}/E_{2g}$ peak intensity ratio vs. the $E_{2g}$ frequency when the data set is restricted to the range of $A_{1g}$ and $E_{2g}$ linewidths indicated in the inset.*

As previously mentioned, $I_A/I_E$ is influenced by the Raman laser polarization in addition to the grain orientation. This is because the light emitted by these two modes is polarized (linear for $A_{1g}$, elliptical for $E_{2g}$)[55], causing $I_A/I_E$ ratios to vary based on the orientation of the grains with respect to incident laser light as well as the Raman setup itself. The latter effect can be mitigated by using polarized Raman spectroscopy, which makes use of a polarizer to only analyze components of the Raman signal that are crossed (or parallel) to the polarization of the Raman excitation laser. With a cross-polarized setup, the $A_{1g}$ signal is suppressed for crystallites oriented in-plane. Consequently, small values of $I_A/I_E$ indicate that the crystal grains are oriented in-plane, while larger values of $I_A/I_E$ indicate that the crystal grains are oriented at an angle relative to the crystal plane. Polarized Raman data was therefore collected to further investigate the trends revealed by Figures 2(a-b). This was done in the same fashion as in the unpolarized case, but using a single film thickness for simplicity, resulting in over 900 unique polarized Raman spectra. For this purpose, the 3.2 nm thick film was used, as it showed the clearest linear trend (Figure S6). For each Raman scan across the width of each line, both cross- and parallel-polarized configurations were used (commonly denoted as zXYz and zXXz, respectively). The resulting $\omega_E$ and $\Gamma_E$ values for the cross-polarized case are plotted in Figure 2(c) with the color-scale indicating the cross-polarized $I_A/I_E$ ratio. A 3D plot of all three variables is also shown in Figure 2(d) with 2D projections along each axis and the color-scale indicating the linear power density. These variables serve as proxies to the structural properties of the lattice, namely: grain size (higher $\Gamma_E$, smaller grain size), strain (higher $\omega_E$, less tension), and grain orientation (higher $I_A/I_E$, larger angle relative to the crystal plane). It is worth emphasizing here that, as the spatial resolution of the Raman is 1 μm, the measured values constitute averages over many crystal grains. Overall, higher $I_A/I_E$ ratios are observed at higher $\omega_E$ values and lower $\Gamma_E$ values. This suggests that out-of-plane oriented (OoPO) grains tend to be larger (lower $\Gamma_E$) and under less strain (higher $\omega_E$). It is worth noting that larger grains are still able to form with in-plane orientation, as can be seen in the larger spread in Figure 2(e) at low $\Gamma_E$ values, while smaller grains tend to be almost exclusively in-plane. While it is difficult to causally link any two variables, it is likely that OoPO grains are forming due to the higher annealing temperatures, which also naturally tends to result in larger grains and less strain, as discussed earlier. This is supported by Figure 2(e), which shows a violin plot of $I_A/I_E$ as a function of the linear power density. Lower annealing temperatures tend to result in lower ratios with a smaller variance, corresponding to a large number of in-plane-oriented (IPO) grains. In contrast, higher annealing temperatures show larger ratios with a larger variance, corresponding to increased populations of both OoPO and IPO grains. It is also possible that the grain orientation influences the amount of strain in the crystal-plane, with IPO grains experiencing more strain due to lattice mismatch from neighboring grains on all sides. This can be evaluated by restricting the values of $\Gamma_A$ and $\Gamma_E$ to a small range of values (determined by binning the data and selecting the most-populated bin), resulting in the plot shown in Figure 2(f). Despite the restricted linewidth values, a positive correlation between $I_A/I_E$ and the $E_{2g}$ frequency is still observed, supporting the argument that OoPO grains experience less strain.

Another key contribution to sensor performance is the extent of defects within the lattice structure[27,28]. To qualitatively assess this, unpolarized, resonant Raman spectra using a 633 nm excitation laser were collected for the 3.6 nm film. When excited with an excitation wavelength close to the direct bandgap of $MoS_2$, non-zone center and combination modes emerge that are not observed in non-resonant Raman spectra (i.e. with 514.5 nm excitation). The emergence of a low frequency longitudinal acoustic (LA) mode centered at 225 cm$^{-1}$ corresponds to scattering of LA phonons at the M point in the Brillouin zone and has been shown to be directly tied to lattice disorder or defect density[53,59,60]. Example resonant Raman spectra are shown in a waterfall plot in Figure 3(a) where the stacking order and color scale correspond to the LA/$E_{2g}$ peak intensity ratio ($I_{LA}/I_E$). These spectra were each fitted in the same manner as the non-resonant Raman spectra, but with an additional peak to account for the LA mode. As the peaks corresponding to the $A_{1g}$ and $E_{2g}$ modes each contain an additional peak under resonant excitation that have not been accounted for here for simplicity, the extracted fitting parameters for these modes cannot be considered as accurate as those determined from non-resonant excitation. Nevertheless, the general trends between each parameter can still be considered – for example, a higher $I_{LA}/I_E$ ratio will correspond to more lattice disorder for a given crystal size. Plotting $\Gamma_E$ and $\omega_E$ vs. $I_{LA}/I_E$ (Figure 3(b-c)) shows a clear trend where regions with smaller grains (higher $\Gamma_E$) and more strain (lower $\omega_E$) tend to have greater defect densities. Coloring each data point based on the translation speed of the sample under laser exposure reveals a clustering based on speed, indicating that the laser processing conditions can directly control defect density. By generating a violin plot of $I_{LA}/I_E$ for each speed (Figure 3(d)), a decreasing trend is observed where regions annealed at slower speeds (i.e. heated for longer times) tend to be more defective. In contrast, higher $P_{lin}$ values show lower $I_{LA}/I_E$ ratios (Figure 3(e)), suggesting that higher annealing temperatures lead to less defective films.

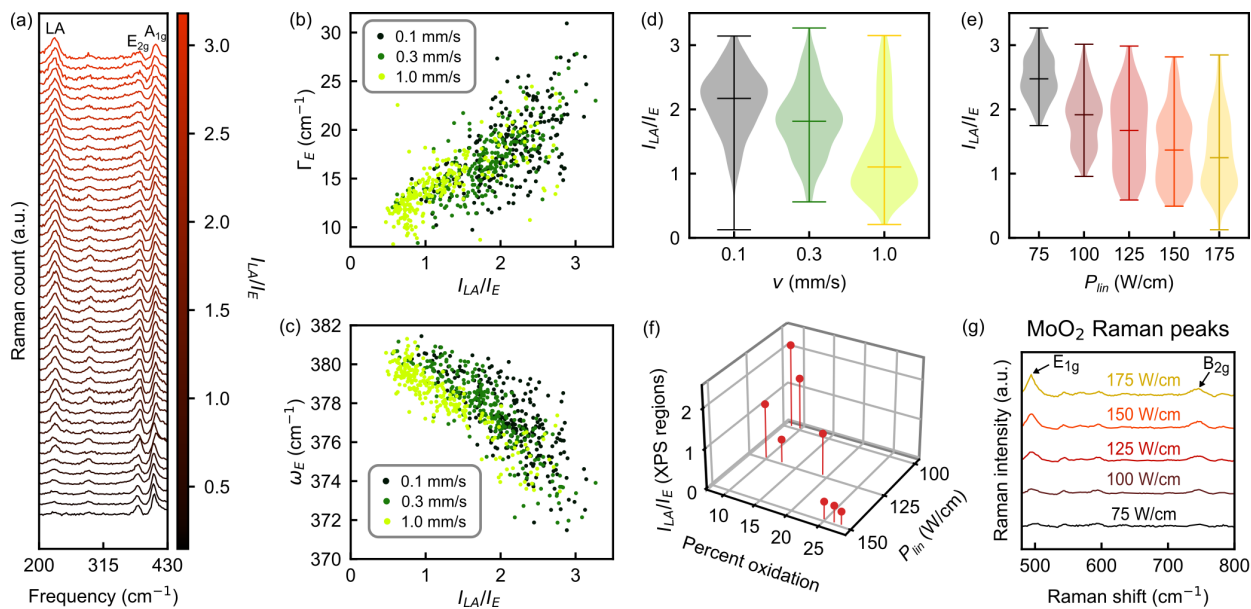

*Figure 3: Analysis of resonant Raman spectra. (a) Waterfall plot of select resonant Raman spectra (3.6 nm film only), stacked and colored based on the LA/E$_{2g}$ intensity ratio. (b-c) Plots of the E$_{2g}$ linewidth and frequency vs. the LA/E$_{2g}$ intensity ratio. The color scale indicates the speed of the translation stage during laser annealing. (d-e) Violin plots of the LA/E$_{2g}$ peak intensity ratio for each v and P$_{lin}$ value. (f) 3D plot of the LA/E$_{2g}$ peak intensity ratio vs. percent oxidation and P$_{lin}$ for 1 mm$^2$ regions formed by raster-scanning the laser at P$_{lin}$ = 100, 125, and 150 W/cm and v = 0.1, 0.3, and 1.0 mm/s. The percent oxidation was determined by acquiring and fitting XPS spectra of each region and extracting the relative intensity of the Mo 3d(+6, Mo$_x$O$_y$) peak. (g) Non-resonant Raman spectra for v =1 mm/s regions showing E1g and B2g peaks characteristic of MoO$_2$. Each spectra represents the median value of all spectra across the width each raster-scanned region, and is normalized to the intensity of the 2H-MoS$_2$ A$_{1g}$ peak.*

Raman spectroscopy alone, however, is not sufficient to determine the type of defects present. In particular, it is expected that the laser-annealing process could potentially result in oxidation due to the presence of residual oxygen in the argon environment. By performing x-ray photoelectron spectroscopy (XPS), the degree of oxidation was determined. This was performed on 1 mm$^2$ regions (corresponding to twice the XPS spot size) formed on the same 3.6 nm sample used previously by raster-scanning the laser at various critical conditions determined through initial screening of the Raman data above. The percent oxidation was determined by fitting the resulting spectra and extracting the intensities of the relevant peaks (see SI for details). Before undergoing XPS analysis, the $I_{LA}/I_E$ ratio of each region was characterized with resonant Raman spectroscopy in the same manner as the previous laser-annealed regions. Figure 3(f) shows the resulting 3D plot of the $I_{LA}/I_E$ ratio vs. the percent oxidation and $P_{lin}$, where this ratio is found to decrease with the degree of oxidation. The $I_{LA}/I_E$ ratio for these XPS regions is also found to decrease with increasing values of $P_{lin}$, consistent with the results from the larger data set in Figure 3(e). Altogether, higher annealing temperatures are found to simultaneously result in both more oxygen and fewer defects. This suggests that the oxygen present in the regions annealed at lower temperatures acts as a substitutional defect, while much of the oxygen present in the regions annealed at higher temperatures does not. It is proposed here that, at higher annealing temperatures, the MoS2 reacts with oxygen to form $Mo_xO_y$ crystallites. Oxygen atoms within these crystallites no longer act as defects within the MoS2 lattice, and the $I_{LA}/I_E$ is consequently found to decrease. This also explains the relationship observed in Figure 1(e) between the carrier density and $P_{lin}$: regions that were annealed at lower temperatures show a higher degree of p-doping due to the presence of oxygen dopants. However, as $P_{lin}$ increases, the density of oxygen dopants decreases as they become locked away in $Mo_xO_y$ crystallite bonds, resulting in regions that are less p-doped. This hypothesis can be confirmed by analyzing the non-resonant Raman spectra at higher frequencies where characteristic $Mo_xO_y$ peaks can be found. Figure 3(g) shows the spectra of the 3.6 nm sample for each value of $P_{lin}$ with the scan speed fixed at v =1 mm/s. These spectra are the median values across the width of each raster-scanned region, normalized to the intensity of the 2H-MoS2 $A_{1g}$ peak. Peaks characteristic of $MoO_2$ are apparent, most notably the $E_{1g}$ and $B_{2g}$ peaks. These peaks are found to increase with $P_{lin}$, indicating that $MoO_2$ crystallites are forming more readily at higher annealing temperatures. These results also offer a potential explanation for the decrease in $I_{LA}/I_E$ ratio with increasing scan speed: as observed in a previous study of laser-annealed 2H-MoS2[47], $MoO_2$ is a reaction intermediate that forms most readily at higher scan speeds (shorter heating times) when formed through laser-annealing, and begins oxidizing further at slower scan speeds (longer heating times). However, it is not clear why crystalline $MoO_3$ does not form at these scan speeds (no characteristic $MoO_3$ Raman peaks are observed for any region); further studies covering few-layer 2H-MoS2 properties over a wider range of scan speeds are needed.

Finally, it is also worth noting that the majority of oxygen dopants present in the regions annealed at lower temperatures were not introduced during the laser-annealing process. This is evidenced by the 11% oxidation measured for the precursor film, which is comparable to the 9-12% oxidation measured for regions annealed at $P_{lin}$ = 100 W/cm. Instead of introducing additional oxygen, lower annealing temperatures primarily redistribute extant oxygen atoms into a crystalline structure.

**High-throughput evaluation of electronic and optical characteristics**

Further characteristic information about each laser-annealed region can be determined by complementary optical and electrical measurements. For optical absorbance measurements, the sample with the largest pre-annealed film thickness (7.8 nm) was chosen as it exhibits the greatest optical density for intensity-dependent optical characterization. Representative absorbance spectra are shown in Figure 4(a) as a function of increasing power involving laser crystallization. Note that the measured absorbance spectra constitute ensemble responses for each annealed region (spatial resolution limit of approximately 55 μm²), rather than capturing the variation within each region as in the case of Raman spectroscopy. Although this limits the number of spatially resolved optical data points, a comparison to the Raman data was performed by calculating the median value of each extracted Raman spectra parameter (i.e., $\Gamma_{E,A}$, $\omega_{E,A}$, $I_{E,A}$, $\epsilon$, and $n$). In this way, the ensemble optical intensity responses can be correlated to the higher resolution Raman characteristics.

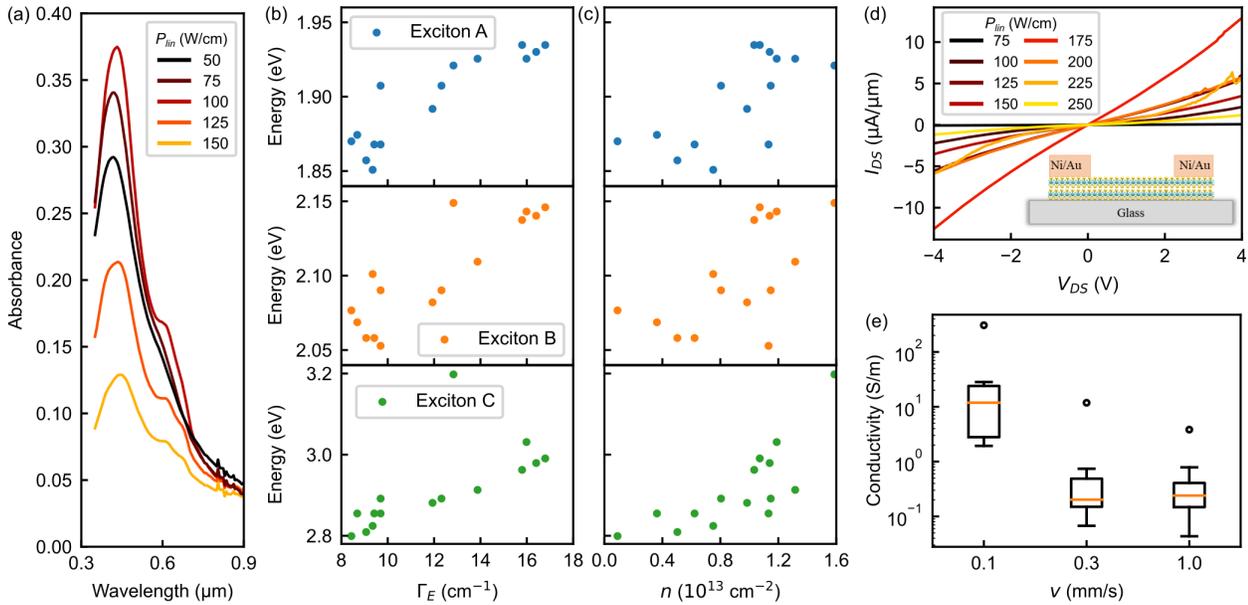

*Figure 4: Optical and electrical characterization. (a) Absorbance spectra for select regions of the 7.8 nm film. (b) Exciton energies vs. $E_{2g}$ FWHM linewidth. (c) Exciton energies vs. the Raman-derived carrier concentration. (d) IV-curves for each select regions of the 3.6 nm film. Inset: diagram illustrating the geometry of the measurement setup. (e) Boxplot of the resulting conductivity vs. scan speed.*

From the peaks in the absorbance spectra, the associated A, B, and C exciton energy shifts for $MoS_2$ can be related to predicted or known structural characteristics. For prototypical semiconducting 2H-$MoS_2$, the A and B excitons correspond to the energy gap between the conduction band minimum and the two valence band maxima associated with the spin-orbit split valence bands[61]. Additionally, the C exciton corresponds to a region in the band structure where the valence and conduction bands are nearly parallel, leading to band-nesting and strong absorption[61]. Characterizing peak exciton energies can subsequently yield information about the band structure within the 2H-$MoS_2$ laser-crystallized region of interest. Across the laser-crystallized regions, a blue-shift in the exciton energies is observed (~0.08 eV for the A and B excitons and ~0.19 eV for the C exciton) relative to the $E_{2g}$ linewidth (Figure [4](b)). This suggests that regions with smaller grains tend to have a wider band gap between the valence and conduction bands, particularly near the band-nesting region. Wendumu *et al.*[62] theoretically investigated the influence of nanocluster (i.e., <10 nm) $MoS_2$ lateral grain sizes and show there is an intrinsic interplay between edge metallic properties and inner semiconducting properties as a function of flake (or grain) size. Similar size-dependent spectral features have also been observed for nanocluster $MoS_2$ in solution[63]. Huang *et al.*[11] show that – in the case of monolayer 2H-$MoS_2$ – the strain induced by an increase in the number of grain boundaries (presumably resulting in smaller grain sizes) may lead to a decrease in the band gap. However, the amount of tensile strain investigated by Huang *et al.* (up to ~6%) is significantly greater than that observed here (up to ~1%). The results here instead suggest that the changes in exciton peak energy appear more strongly correlated to changes in the carrier concentration. This is supported by Rao et al.[64] and Stevenson et al.[20] who observed a blue-shift in the energy of the A exciton emission and complex refractive index peak dispersions, respectively, upon passivation of sulfur vacancies with oxygen (i.e. when the material becomes less n-type due to a reduction in the electron density). Plotting the exciton energies vs. the carrier concentration (Figure 4(c)) shows a similar blue-shift when the material becomes more p-type due to the presence of oxygen dopants discussed previously.

Electrical conductivity measurements were also carried out for the film with a 3.6 nm pre-annealed film thickness at various conditions. The resulting IV-curves from select regions are shown in Figure 4(d). Together with the film thickness, the conductivity can be calculated, which represents an averaging over the width of each annealed region. Generating a box plot of the conductivities for each scan speed results in Figure 4(e), showing significantly higher conductivities for the regions annealed at slower speeds. As previously discussed, regions annealed at slower speeds tend to result in a more defective film (Figure 3(b)) due to the presence of oxygen dopants. At higher annealing speeds, the defect density decreases, leading to a corresponding decrease in conductivity.

**Data-driven optimization of thin-film chemical sensors**

Thin films of crystalline $MoS_2$ reveal significant promise in electronic chemical sensing devices[21,65,66], where modulations of the carrier density through binding of a molecule to the surface modifies the electrical conductivity within the channel. As the laser-crystallization approach described in this work highlights access to a broad range of grain sizes, defect densities, strain, carrier concentrations, and grain orientations, chemical sensors were fabricated using selected conditions to evaluate their relationship to sensitivity and limit of detection. For this second part of the study, a subset of these material properties were realized on samples with film thicknesses of 2.4, 3.6, and 7.8 nm. These sensor devices were made by first fabricating multiplexed Cr/Au contacts onto a glass substrate with a channel gap with dimensions of 200 μm². Amorphous $MoS_2$ was then deposited over the channel gaps for each device by using a mask with 1.4 mm by 0.9 mm rectangular holes. The channel gaps for each sample were annealed by raster-scanning the laser with different linear power densities and scan speeds (Figure 5(a)), producing sensing regions with a wide range of material properties. The sensing regions were then characterized by acquiring multiple Raman spectra over the width of each region using unpolarized, polarized, and resonant Raman. As was done previously, the peak intensities, linewidths, and frequencies were determined by fitting the spectra, which could then be used as proxies to material properties of interest. In particular, the $E_{2g}$ linewidth and frequency (from unpolarized Raman), the $LA/E_{2g}$ peak intensity ratio (from resonant Raman), and the $A_{1g}/E_{2g}$ peak intensity ratio (from cross-polarized Raman) were chosen as correlates to grain size, strain, defect density, and grain orientation, respectively. Given that the sensor response is based on the average material properties of the sensing region, the median values of these Raman parameters for each device were used throughout this analysis.

To determine the sensing response of each of these regions, the relative change in resistance ($\Delta R/R$) across 15 unique devices was monitored over time during exposure to 1 ppm of $NO_2$ with the resulting responses shown in Figure 5(b). A strong dependence on the film thickness on overall sensitivity (i.e. the largest value of $\Delta R/R$) was observed, and is further illustrated in the boxplot in Figure 5(c). Thinner films consistently show a stronger response; this is expected given that a higher surface area to volume ratio leads to a larger concentration of surface charge carriers reactive to the influence of adsorbed $NO_2$ molecules compared to the bulk charge carrier concentration, resulting in a greater change in conductivity. While the variance of the devices with a 2.4 nm thick film indicates that other material properties influence sensitivity, it is clear that film thickness is the most important factor.

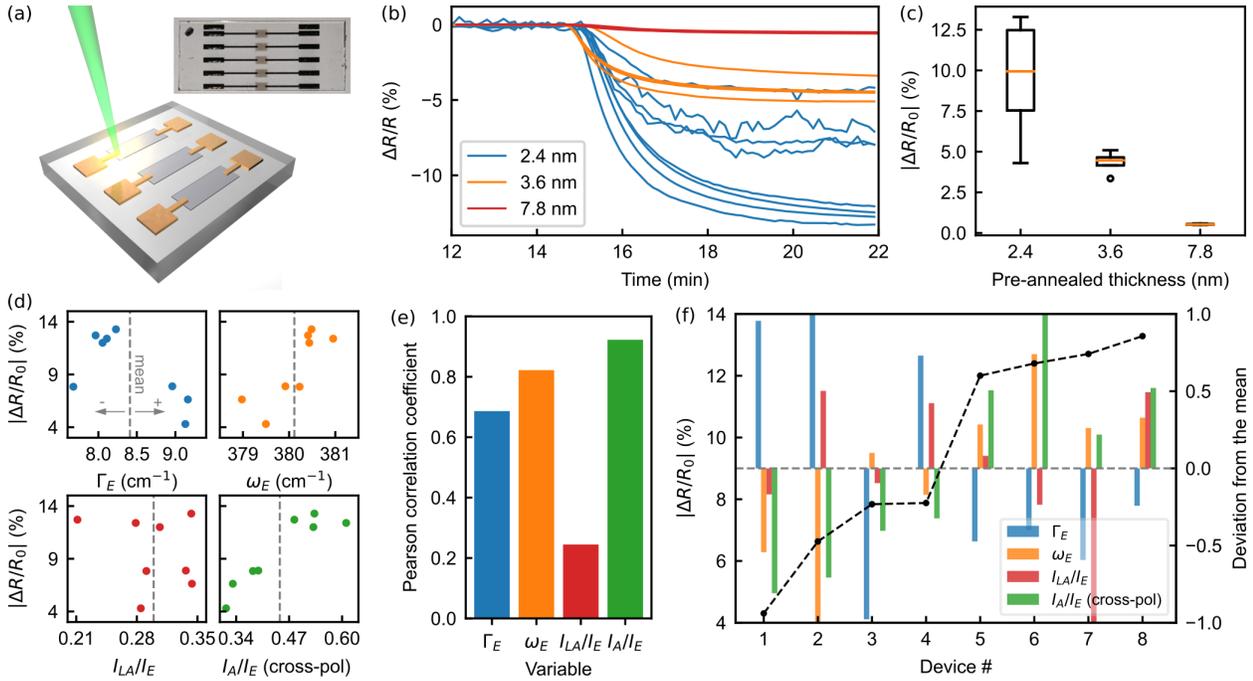

Figure 5: Gas sensor testing and analysis. (a) Schematic illustrating the geometry of the sensors: a sensing region (laser-annealed 2H-MoS$_2$) between two gold contacts. Inset: image of a sensor with the a-MoS$_2$ film clearly visible between contacts. (b) Plot of the sensor response (relative change in resistance) to 1 ppm of $NO_2$ for each device as a function of time, with the color indicating the film thickness used for the device. (c) Boxplot of the sensor response for each film thickness. (d) Plot of the sensor response vs. the $E_{2g}$ linewidth, $E_{2g}$ frequency, LA/$E_{2g}$ intensity ratio, and the (cross-polarized) $A_{1g}/E_{2g}$ peak intensity ratio. The grey dashed lines correspond to the mean values of each Raman parameter. (e) Bar plot of the Pearson correlation coefficients associated with the plots in (d). (f) Plot of the sensor response of each device overlaid on top of a histogram of the parameters used in (d). The dashed grey line represents the mean value of each Raman parameter, and the height of the bars represent the deviation from the mean relative to the maximum deviation.

For the thinnest (2.4 nm) film, a total of 8 devices were created, each under different annealing conditions. The effects of each Raman parameter on sensor response are shown in Figure 5(d) where the $\Delta R/R$ of each device is plotted as a function of each Raman parameter. The results are summarized in a bar plot of the Pearson correlation coefficient (the ratio between the covariance and the product of the standard deviations) in Figure 5(e). When comparing the Raman parameters of these devices, the $I_A/I_E$ ratio from the cross-polarized Raman spectra showed the strongest correlation with sensor response. As this ratio corresponds to the orientation of crystal grains, this suggests that OoPO grains tend to result in a greater change in device conductivity, consistent with the results reported by Cho *et al.*[26]. The effects of the Raman parameters are further illustrated in Figure 5(e) where the sensor response is plotted for each device alongside a histogram representing the deviation of each variable from its mean value (dashed grey lines in Figures 5(d,f)). Bars with negative deviation correspond to devices where the associated Raman parameter fell below the mean value depicted by the dashed grey line in the corresponding plot in Figure 5(d). Similarly, bars with positive deviation correspond to devices where the associated Raman parameter fell above the mean value. In addition to the $I_A/I_E$ ratio, correlations with the $E_{2g}$ linewidth and frequency are also present to a lesser extent, which is to be expected as these variables correlate with the $I_A/I_E$ ratio itself, as discussed earlier. However, it is also possible that strain directly influences sensor response – further studies would be needed to distinguish between the effects of strain and grain orientation. Lastly, it is interesting to note that no clear correlation with the $I_{LA}/I_E$ ratio is seen, suggesting that the presence of defects does not play a significant role in sensor response relative to the other variables for the range of defect densities measured here. This is likely due to rapid termination of highly reactive surface sites by ambient species such as oxygen.

## Conclusions

In order to create tailored devices built around 2D materials, strategies that enable high-throughput, data-driven design will become increasingly important. In this work, a laser-crystallization strategy was used to create hundreds of annealed regions on various samples that reveal unique structure/property relationships. Through non-resonant Raman analysis of over 4000 unique Raman spectra, a clear linear dependence upon the linewidth of in-plane and out-of-plane vibrational modes was observed as well as a linear interplay between strain and doping. The large library of spectroscopic data enabled the use of a UMAP data visualization approach, revealing clustering of $I_A/I_E$ and encouraging further investigation. To quantitatively evaluate this ratio, polarized Raman was employed to explore the role of orientation on strain, doping, and crystal size in laser-annealed $MoS_2$ films. Resonant Raman spectroscopy enabled the high-throughput analysis of lattice disorder and strategies to induce disorder in materials through controlling the annealing time (i.e. scan speed in the laser-processing setup). Optical, electronic, and sensing properties of the laser-crystallized $MoS_2$ thin films were compared to the repository of resonant and non-resonant Raman spectra. The ability to induce high-throughput, specific, controllable intrinsic properties at this scale was used to repeatably fabricate large-scale conductometric chemical sensors from laser-crystallized $MoS_2$ regions on substrates with patterned metal contacts. Film thickness, grain size, and grain orientation were shown to be the primary contributors to sensitivity in 2D $MoS_2$ chemical sensors fabricated in this study, with film thickness showing the greatest influence. Collectively, the results of this study demonstrate how high-throughput data generation of 2D material processing conditions can accelerate the discovery of material processing mechanism, as well as the optimization of functional devices. For example, recreating this data set using conventional chemical vapor deposition methods would have increased the timeline of this study by a factor of ~1000 (assuming ~1 hour vs. ~1 second per processing candidate). Affordable generation of 2D material processing training data sets that will also empower the development of machine learning surrogate models for integration into the fabrication workflow[67]. Using convolutional neural networks or other neural network techniques, optical images of the laser-annealed film can potentially be mapped to material-level or device-level performance characteristics. This mapping will not only accelerate the calibration of processing conditions during early stage material development, but could also serve as a rapid assessment tool for quality control during large scale production. High-throughput, data driven design strategies are an important tool for stimulating these opportunities for the 2D materials community.

# Methods

## Magnetron sputtering

The samples consisted of films of *a*-MoS$_2$ with varying thickness grown on 0.5 mm thick willow glass substrates by magnetron sputtering at a substrate temperature of 25°C. Sputtering was performed via asymmetric bi-polar pulsed direct current magnetron sputtering at 65 kHz (with a 0.4 s reverse time) from a polycrystalline MoS$_2$ target (Plasmaterials) at room temperature with a growth rate of approximately 1 atomic layer every 4 seconds. Growth times of 12, 15, 18, 24, 30, and 36 s were used to produce films with thicknesses of 2.4, 3.2, 3.6, 4.9, 6.5, and 7.8 nm, respectively.

## Laser processing

The laser processing setup used is identical to that described by Austin *et al.*[47], but with a 100x magnification, infinity-corrected objective (Mitutoyo N02027112) with a numerical aperture of 0.55. The $1/e^2$ beam radius during processing was measured to be $10.0 \pm 0.2$ μm. Each sample was placed in an environmental gas cell and brought to rough vacuum (30 mTorr) before 99.999% purity argon gas was flowed into the cell with an equilibrium pressure of 11 Torr before laser-processing. Each laser-written region was formed by raster-scanning the laser eight times with a separation of 6 μm between each raster and with a length of 1 mm.

## Raman spectroscopy

The Raman spectra were obtained using a Renishaw InVia with 1800 and 1200 lines/mm gratings for the 514.5 and 633 nm excitation lasers, respectively. A 100x objective lens with a numerical aperture of 0.85 was used, resulting in a spatial resolution of 1 μm. The laser power was kept below 1 mW to minimize any heating by the laser. Spectra were collected from unannealed *a*-MoS$_2$ and select laser-crystallized 2H-MoS$_2$ regions over the course of 10 consecutive 30 second acquisitions with 2 accumulations each.

## TEM specimen preparation and electron microscopy

The TEM specimens of the films were prepared via focused ion beam (FIB) milling process employing a Helios Nano Lab 660 FIB unit. STEM images were recorded on a Nion UltraSTEM 100 operating at 100 kV, with a convergence angle of 32 mrad. The HAADF detector possessed an inner collection angle of ~80 mrad and an outer collection angle of ~200 mrad. The STEM was equipped with a Gatan Enfina electron energy-loss (EEL) spectrometer and the EELS experiments were performed with a convergence semi-angle of 30 mrad, and an EELS collection semi-angle of 48 mrad.

## Ultraviolet-visible spectroscopy

Absorption spectra were collected using a CRAIC UV-vis microspectrophotometer via transmission using a 15x objective over a wavelength range of 300–1600 nm. Effects of dark current and the substrate were accounted for by taking a dark scan and a reference scan of the substrate.

**Conductivity measurements**

In order to perform conductivity measurements, resistors were fabricated using two e-beam lithography steps. In both steps, the glass substrate was spin coated with MMA and PMMA photoresist stacks. To avoid charging effects associated with glass (insulating) substrate, 20 nm of Au is deposited on top of the MMA/PMMA stack. Following that, e-beam lithography was used to expose the desired regions. After exposure, Au was removed using trifluoroacetic acid and the exposed regions were developed using MIBK and IPA. In the first step, the $MoS_2$ channel was defined, and $MoS_2$ was etched from the rest of the sample using SF6. In the second step, e-beam evaporation was performed to deposit 40 nm Ni/30 nm Au stack. Finally, the photoresist is removed using acetone and IPA in both cases. Conductivity measurements were performed using a Keysight B1500A parameter analyzer inside a Lake Shore CRX-VF probe station at room temperature in high vacuum (~$10^{-6}$ Torr).

**Sensor testing**

Sensing was tested in a homebuilt 2.2 L stainless steel chamber. The sample was placed inside and flushed overnight with dry oxygen. Resistance measurements were taken with a Keysight DAQ970A containing a DAQM900A 20-channel solid-state multiplexer module at 6.5 digits of precision integrating over 10 power line cycles with connection fed through a 37-pin subminiature-D CF flange (Kurt Lesker). A mixture of and 10 ppm $NO_2$ balanced with nitrogen (Indiana Oxygen) was mixed with dry nitrogen using mass flow controllers (MKS Instruments 1179C series) to produce the desired concentration (1:9 for 1 ppm). Total flow was held at 2000 sccm.

## Conflict of interest

The authors declare no competing financial interest.

## Acknowledgments

N. G. and D. A. acknowledge support from the Asian Office of Aerospace Research and Development within the Air Force Office of Scientific Research grant #21IOA110 as well as support from the Air Force Office of Scientific Research under grant number 20RXCOR057. S. K. and G. B. acknowledge support from the DAGSI fellowship. STEM experiments conducted as part of a user proposal at the Center for Nanophase Materials Sciences, which is a DOE Office of Science User Facility.

STEM experiments conducted as part of a user proposal at the Center for Nanophase Materials Sciences, which is a DOE Office of Science User Facility.

## Supporting information available

Additional experimental and analytical details can be found in the supporting information. This includes full-size microscope and STEM images, a more-detailed description of the Raman spectra analysis, and additional plots of variables derived from the Raman spectra.